\documentclass{aa}

\usepackage{natbib}
\usepackage{amsmath}
\bibpunct{(}{)}{;}{a}{}{,}
\usepackage{graphicx}
\usepackage{txfonts}

\begin{document}
\title{Photochemistry of the PAH pyrene in water ice: the case for ion-mediated solid-state astrochemistry }
\titlerunning{PAH ice photochemistry}


\newcommand{\water}{H$_2$O }
\newcommand{\co}{CO$_2$ }
\newcommand{\wav}{cm$^{-1}$ }

\author{Jordy Bouwman \inst{1}
          \and Herma M. Cuppen \inst{1} 
          \and Arthur Bakker \inst{1}          
			\and Louis J. Allamandola  \inst{2}
          \and Harold Linnartz \inst{1}
           }

   \offprints{Jordy Bouwman, bouwman@strw.leidenuniv.nl}

  \institute{Raymond and Beverly Sackler Laboratory for Astrophysics, Leiden Observatory, Leiden University, P.O. Box 9513, NL 2300 RA Leiden, The Netherlands
\and NASA-Ames Research Center, Space Science Division, Mail Stop 245-6,
Moffett Field, CA 94035}

\date{Received ; accepted}


  \abstract
   {Icy dust grains play an important role in the formation of complex inter- and circumstellar molecules. Over the past decades, laboratory studies have mainly focussed on the physical interactions and chemical pathways in ices containing rather simple molecules, such as H$_2$O, CO, CO$_2$, CH$_4$ and CH$_3$OH. Observational studies show that polycyclic aromatic hydrocarbons (PAHs) are also abundantly present in the ISM in the gas phase. It is likely that these non-volatile species freeze out onto dust grains as well and participate in the astrochemical solid-state network, but experimental PAH ice studies are largely lacking.}
   {The study presented here focuses on a rather small PAH, pyrene (C$_{16}$H$_{10}$), and aims to understand and quantify photochemical reactions of PAHs in interstellar ices upon vacuum ultraviolet (VUV) irradiation as a function of astronomically relevant parameters.}
	{Near UV/VIS spectroscopy is used to track the \itshape in situ \upshape VUV driven photochemistry of pyrene containing ices at temperatures ranging from 10 to 125~K.}
	{The main photoproducts of VUV photolyzed pyrene ices are spectroscopically identified and their band positions are listed for two host ices, \water and CO. Pyrene ionisation is found to be most efficient in \water ices at low temperatures. The reaction products, triplet pyrene and the 1-hydro-1-pyrenyl radical are most efficiently formed in higher temperature water ices and in low temperature CO ice. Formation routes and band strength information of the identified species are discussed. Additionally, the oscillator strengths of Py, Py$^{\cdot+}$ and PyH$^\cdot$ are derived and a quantitative kinetic analysis is performed by fitting a chemical reaction network to the experimental data. } 
{Pyrene is efficiently ionised in water ice at temperatures below 50~K. Hydrogenation reactions dominate the chemistry in  low temperature CO ice with trace amounts of water. The results are put in an astrophysical context by determining the importance of PAH ionisation in a molecular cloud. We conclude that the rate of pyrene ionisation in water ice mantles is comparable to the rate of photodesorption of \water ice. The photoprocessing of a sample PAH in ice described in this manuscript indicates that PAH photoprocessing in the solid state should also be taken into account in astrochemical models. }


\maketitle

\section{Introduction}

Strong infrared emission attributed to polycyclic aromatic hydrocarbons (PAHs) is characteristic of many galactic and extragalactic objects \citep{smith07,draine07,draine07a,tielens08}. While this emission generally originates from optically thin, diffuse regions, PAHs should also be common throughout the dense interstellar medium. There, as with most other interstellar species in molecular clouds, PAHs condense out of the gas onto cold icy grain mantles where they are expected to influence or participate in the chemistry and physics of the ice. While laboratory studies on interstellar ice analogs have shown that complex organic molecules are produced upon extended vacuum ultraviolet (VUV) photolysis \citep[e.g.][]{briggs92,bernstein95}, the photoinduced processes at play during the irradiation of PAH containing interstellar ice analogues have not been studied in detail. In optical, {\itshape in situ} studies on the photochemistry of naphthalene, 4-methylpyrene and quatterylene containing water ice at 20~K, \citet{gudipati03, gudipati06a,gudipati06} and \citet{gudipati04} showed that these PAHs are readily ionized and stabilized within the ice, suggesting that trapped ions may play important, but overlooked roles in cosmic ice processes. Beyond this, there is little information on the VUV induced, {\itshape in situ} photochemistry and photophysics of PAH containing water rich ices.

Here, we describe a detailed study of the VUV induced photochemistry that takes place within pyrene (Py or C$_{16}$H$_{10}$) containing water ices (Py:H$_2$O=1:10,000-1:5,000). The present study is an extension of a recently published study \citep{bouwman09} in which the focus was on the new experimental setup and where the use of PAH ice spectra was discussed to search for solid-state features of PAHs in space. In this work, the focus is on a detailed characterization of the actual chemical processes taking place upon VUV irradiation, particularly as a function of ice temperature ranging from 25 to 125~K. Additionally, measurements on Py:CO ices at 10~K have been performed to further elucidate the role of water in the reaction schemes and to clarify formation routes of identified species. {\bf A similar study of three small PAHs is now underway to understand the general principles of PAH/ice photochemistry.} This is part of an overall experimental program at the Sackler Laboratory for Astrophysics to study the fundamental processes of inter- and circumstellar ice analogues such as thermal \citep{acharyya07} and photodesorption \citep{oberg07a,oberg08a}, hydogenation reactions \citep{fuchs08,ioppolo08}, photochemistry \citep{oberg09a} and physical interactions in interstellar ice analogues \citep{bouwman07,oberg07,oberg09}.

\begin{figure*}[t]
\centering
\includegraphics[width=17cm]{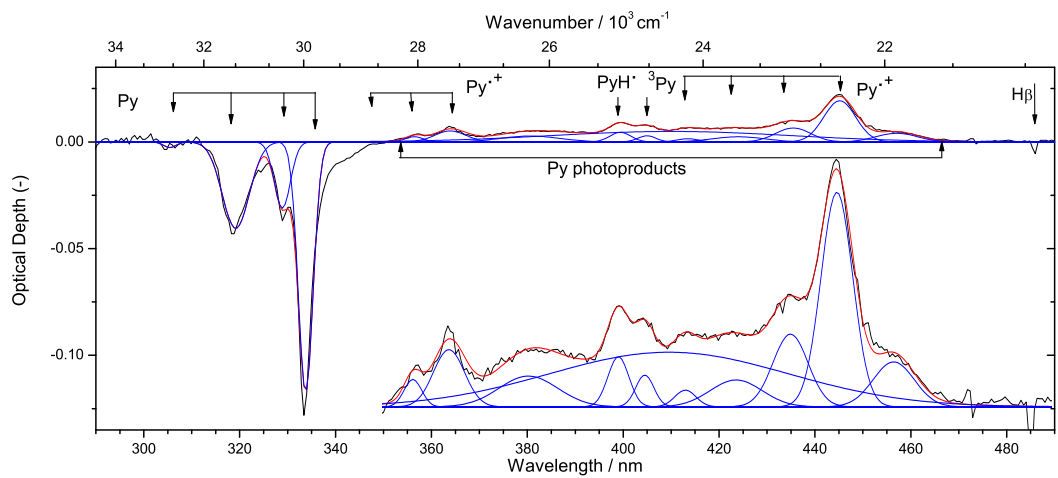}
\caption{The spectrum of a dilute pyrene:H$_2$O ice after 900~s of VUV irradiation at 125~K. The inset shows a blow-up of the pyrene photoproduct bands. Band assignments are discussed in Section~\ref{sectionresults}. Note the broad feature ranging from about 350 to 470~nm which is indicated by a Gaussian fit. This is attributed to overlapping bands from individual pyrene photoproducts. Bands with negative optical depth indicate species destruction, those with positive optical depth show species formation. The blue bands are Gaussian profiles which co-add to the overall fit shown in red. Note the instrumental resolution indicated by the profile of the H$\beta$ line at 486.1~nm. (This figure is available in color in electronic form)}
\label{FIGURE1}
\end{figure*}

The manuscript is organized as follows. The experimental technique is summarized in Section~\ref{experimental}. Section~\ref{sectionresults} describes the Py:H$_2$O and Py:CO ice photochemistry, the resulting products and their formation routes. The temperature dependent photochemistry and derived reaction dynamics are described in Section~\ref{kineticanalysis} and astrochemical implications are discussed in Section~\ref{sectionAstrochemical}. The main conclusions are summarized in Section~\ref{conclusions}.

\section{Experimental Technique}
\label{experimental}

We use a recently constructed apparatus, described in \citet{bouwman09}, that follows the photochemistry in kinetic mode during VUV irradiation by measuring the near UV-Visible absorption spectra of an ice, providing \textquoteleft real-time\textquoteright\ tracking of the reactants and photoproducts. Dilute Py:H$_2$O ice samples ($\sim$1:10,000 - $\sim$1:5,000) and a Py:CO ice sample of comparable concentration are prepared by depositing the vapor from a pyrene sample heated to 40$^{\circ}$C together with \water vapor or CO gas onto a cold MgF$_2$ window. The window is cooled to 10~K in the case of CO deposition or 25~K in the case of \water deposition. The sample window is cooled by a closed cycle He refrigerator. Pyrene (Aldrich, 99\%) and CO (Praxair 99.999\%) are used without further purification. Vapor from water, filtered through a milli-Q purification system and further purified by three freeze-pump-thaw cycles, is used. The sample window is mounted in a high-vacuum chamber ($P\approx 10^{-7}$~mbar). Ice growth rate and thickness are determined with a HeNe laser by monitoring the thin-film interference fringes generated during deposition. Simultaneously, the amount of pyrene is tracked  measuring the integrated strength of the S$_2\leftarrow~$S$_0$ neutral Py transition at 334~nm. Deposition is typically stopped when the optical depth (OD) of Py approaches $\sim$0.15.

The ice samples are photolyzed with the 121.6~nm Ly$\alpha$ (10.6~eV) and the 160~nm molecular hydrogen emission bands (centered around 7.8~eV) generated by a microwave powered discharge in a flowing H$_2$ gas with a vacuum ultraviolet flux of $\sim 10^{15}~{\rm photons~s}^{-1}$ \citep{munoz02}. This results in a photon flux of $\sim 10^{14}$ photons$\cdot$cm$^{-2}$s$^{-1}$ at the sample surface \citep{oberg08a}. 

Absorption spectra of VUV photolyzed Py containing ices are measured with a Xe-arc lamp serving as a white light source. Lenses and diaphragms direct the light through the ice sample along the optical axis determined by the HeNe laser beam after which it is focused onto the entrance slit of a 0.3~m spectrometer. A 150~lines~mm$^{-1}$ grating, blazed at 300~nm, disperses the light onto a sensitive 1024$\times$256 pixel CCD camera with 16~bit digitization. The camera is read out in the vertical binning mode by a data acquisition computer on which the data are converted to absorbance spectra (${\rm OD}=-\ln(I/I_0))$. This configuration spans the 270 to 830~nm spectral range, which permits simultaneously monitoring the behavior of the neutral Py parent molecule and photoproduct bands without adjusting any element along the optical path. This is critical to obtain reliable and reproducible baselines in measuring the optical spectra of ices. The spectral resolution is of the order of 0.9~nm which is more than sufficient to record broad solid-state absorption features.

The measurements described here were made on various H$_2$O:Py ice samples at 25, 50, 75, 100 and 125~K. The CO ice experiments were carried out at 10~K to avoid matrix sublimation at higher temperatures. The sample temperature is maintained using a resistive heater with an accuracy of $\pm 2~{\rm K}$. The measured spectra are converted into units of optical depth using the spectrum of the freshly deposited, unphotolyzed ice at the appropriate temperature as a reference spectrum ($I_0$). Recording a single spectrum typically takes about 5~ms and 229 spectra are generally co-added to improve the S/N of a spectrum, resulting in one complete spectrum every 10 seconds. 

The optical configuration of the apparatus is such that spectra are recorded {\itshape simultaneously} with photolysis. Thus, the fast spectral recording time permits monitoring photoinduced changes on a roughly 10 second time scale. Figure~\ref{FIGURE1} shows the 290 to 490~nm spectrum of a Py:H$_2$O ice at 125~K after 900~s of {\itshape in situ} VUV photolysis. Because the spectrum recorded before VUV irradiation is taken as a reference ($I_0$), bands with positive OD values originate from species that are produced by photolysis while the bands with negative OD correspond to the neutral pyrene that is lost upon photolysis. Comparing the Py and photoproduct absorption bands with the narrow H$\beta$ lamp line at 486.1~nm shows that the instrumental resolution indeed far exceeds the ice band widths. The absolute wavelength calibration is accurate to within $\pm$0.5~nm.

More than 1400 individual spectra are recorded and are reduced in a typical 4~hr experiment. Spectra are individually baseline corrected by fitting a second order polynomial through data points where no absorptions occur and subsequently subtracting the fit from the measured spectrum. Integrated absorbances of absorption features are calculated numerically for all spectra. These are corrected for contributions by atomic hydrogen lines originating from the H$_2$ discharge lamp. The data reduction software also allows plotting of correlation diagrams between integrated absorbances of different absorption features. All data handling and reduction is done with LabView routines.

Integrated band areas are used, in conjunction with oscillator strengths ($f$), to derive molecular abundances. The oscillator strength is converted to integrated absorbance (cm~molecule$^{-1}$) using the conversion factor 8.88$\times 10^{-13}$ \citep{kjaergaard00}. The number of molecules per cm$^2$ ($N$) is given by:
\begin{equation}
N=\frac{\int_{\nu_1}^{\nu_2} \tau\, d\nu}{8.88\times 10^{-13}f},
\label{eqns1}
\end{equation}
\noindent where $\tau$ is the optical depth and $\nu$ is the frequency in cm$^{-1}$.

\section{Band assignments and band strength analysis}
\label{sectionresults}

The typical photolysis duration of about 4 hours is the time required for nearly complete loss of the neutral pyrene vibrational progression at 334.0, 329.2 and 319.2~nm. Irradiating the sample ices with VUV light produces a set of new absorption bands in the spectra, indicating active photochemistry. The band positions, FWHM and assignments of the bands in the Py:\water\ ice at 25~K are listed in Table~\ref{table1}. The bands appearing in the Py:CO ice at 10~K are similar to those in the Py:\water ice at 25~K, however, with slightly altered band positions and FWHM and with very different relative intensities (see also Table~\ref{table1}). Figure~\ref{FIGURE1} presents a spectrum from the 125~K Py:H$_2$O series. This figure illustrates production of the pyrene radical cation (Py$^{\cdot +}$), triplet pyrene ($^3$Py), 1-hydro-1-pyrenyl radical (PyH$^{\cdot}$) and a broad underlying \textquoteleft residue\textquoteright\ feature upon VUV irradiation. Additionally, a progression of distinct absorptions is found in the Py:CO experiment, which indicates the formation of the (reactive intermediate) HCO$^\cdot$ radical. The identifications of these species and their oscillator strengths are discussed below.

\begin{table}[ht]
\begin{minipage}{\columnwidth}
\caption{Band positions ($\lambda_c$) and FWHM of Gaussian fits to the observed features of neutral pyrene for pure pyrene ice at 10~K, pyrene in a water ice at 25~K and pyrene in a CO ice at 10~K. Photoproduct bands arising from VUV processing of the ices are also listed for the Py:\water and Py:CO experiments. Both band positions and FWHMs are indicated in nm. } \label{table1} \centering
\scriptsize
\begin{tabular}{l c c c c c c}
\hline\hline
 Species&\multicolumn{2}{c}{Pyrene}&\multicolumn{2}{c}{Pyrene:H$_2$O}&\multicolumn{2}{c}{Pyrene:CO}\\
&$\lambda_c$&FWHM&$\lambda_c$&FWHM&$\lambda_c$&FWHM\\
\hline
Py	$^1B_{2u}$				&312.7	&7.1		&319.2	&6.5		&319.4	&7.5	\\
					&325.3	&10.0		&329.2	&3.2		&329.2	&2.3	\\
					&341.5	&14.0		&334.0	&4.4		&334.3	&4.1	\\
Py$^{\cdot +}$ $^2B_{1u}$	&...&...				&363.2	&3.6		&...$^a$	&...$^a$		\\
					&...&...				&354.0	&6.5		&...$^a$	&...$^a$		\\
					&...&...				&344.9	&6.2		&...$^a$	&...$^a$		\\
Py$^{\cdot +}$ $^2A_u$	&...&...				&445.6	&6.6		&445.3	&7.8			\\
					&...&...				&435.5	&10.2		&...$^a$	&...$^a$			\\
					&...&...				&423.0	&12.2		&...$^a$	&...$^a$		\\
					&...&...				&413.8	&5.3		&...$^a$	&...$^a$		\\
Py$^{\cdot +}$	$^2B_{1u}$&...&...				&490.1	&10.0		&...$^a$	&...$^a$			\\
PyH$^\cdot$		&...&...				&399.4$^b$	&5.2$^b$		&400.5	&4.2		\\
					&...&...				&...$^a$	&...$^a$	&392.5	&6.7			\\
					&...&...				&...$^a$	&...$^a$	&378.4	&15.7			\\
$^3$Py $^3A_{g}^-$		&...&...				&405.0$^b$&4.5$^b$&406.2	&4.8			\\
HCO$^\cdot$ $^2A$''		&...&...				&...&...				&513.4		&17.5\\
					&...&...				&...&...				&535.3		&12.5\\
					&...&...				&...&...				&556.3		&14.5\\
					&...&...				&...&...				&583.0		&16.8\\
					&...&...				&...&...				&604.9		&10.0\\
					&...&...				&...&...				&639.2		&15.1\\
\hline
\end{tabular}
\end{minipage}\\
{\scriptsize $^a$	Absorption feature was too weak to perform an accurate fit\\
$^b$	Features are too weak at 25~K; the 125~K values are indicated\\}
\end{table}

\subsection{Neutral pyrene bands}

As in \citet{bouwman09}, the strong, negative bands peaking at 334.0~nm and weaker bands at 329.2 and 319.2~nm in the \water ice (see Fig.~\ref{FIGURE1}), and at slightly shifted positions in the CO ice, are assigned to the $^1B_{2u}$$\leftarrow$$^1A_g$ electronic transition of neutral pyrene (S$_2\leftarrow$~S$_0$) based on previous studies of pyrene in rare gas matrices \citep{vala94,halasinski05}. To study the chemistry in absolute number densities, a value of $f=0.33$ is adopted from the literature for the oscillator strength of pyrene \citep{bito00,wang03}. This value is used throughout this paper both for the Py:\water and Py:CO experiments. Pure pyrene ice measured at 10~K shows broader absorptions located at 341.5, 325.3, and 312.7~nm (see Table~\ref{table1}). We did not perform VUV experiments on the pure pyrene sample.

\subsection{Pyrene cation bands}
\label{cationabsorption}

Positive bands at 363.2, 354.0 and 344.9~nm appear upon photolysis in the Py:\water experiments. This progression is assigned to the $^2B_{1u}$$\leftarrow$$^2B_{3g}$ vibronic transition of the pyrene cation (Py$^{\cdot+}$) in accordance with the proximity to the band positions reported by \citet{vala94} and \citet{halasinski05}. This transition for Py$^{\cdot+}$ in \water\ ice was reported in \citet{bouwman09}. The $^2B_{1u}$$\leftarrow$$^2B_{3g}$ transition is too weak to be detected in the Py:CO experiment. A stronger Py$^{\cdot+}$ progression falls at 445.6, 435.5, 423.0, and 413.8~nm in water ice. Of these bands only the strongest at 445.3~nm is detectable in the irradiated Py:CO ice. This progression is assigned to the $^2A_u$$\leftarrow$$^2B_{3g}$ of Py$^{\cdot+}$. The much weaker absorption due to the $^2B_{1u}$$\leftarrow$$^2B_{3g}$ Py$^{\cdot+}$ transition at 490.1~nm in \water is again undetectable in CO. 

Py$^{\cdot+}$ formation in these \water and CO ice experiments is the result of direct single photon ionisation of the neutral species, following:
\begin{equation}
{\rm Py}~\xrightarrow{VUV}~{\rm Py^{\cdot+} + e^-}.
\label{eq1}
\end{equation}
It has to be emphasized that ionisation in Py:\water ices is far more efficient than in Py:CO ices. Additional measurements on Py:CO:\water mixtures indicate that the presence of \water indeed enhances the ionisation. Hence, it is possible that water contamination in the CO ice is responsible for the formation of some, if not all, of the cation species in the Py:CO experiment. The role of water contamination in CO ice will be discussed in more detail in Section~\ref{sectionHCO}.

\begin{figure}[t]
\centering
\includegraphics[width=8.5cm]{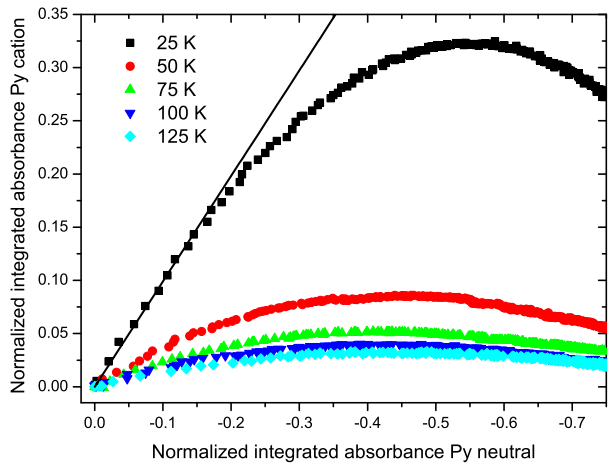}
\caption{Integrated absorbance of the 445.6~nm Py$^{\cdot+}$ band growth plotted against the loss of the 334.0~nm Py band in the 25, 50, 75, 100, and 125~K ices. 10 seconds elapse between subsequent data points. Values are normalized to the maximum integrated absorbance of neutral pyrene. The straight line portion of these plots is used to determine the oscillator strength of Py$^{\cdot+}$ as described in Section~\ref{cationabsorption}.}
\label{FIGURE2}
\end{figure}

Using baseline corrected spectra as shown in Fig.~\ref{FIGURE1}, the photochemical evolution is tracked by integrating areas of bands due to each species in every spectrum and plotting them as a function of photolysis time. The strongest Py$^{\cdot+}$ band at $\sim$445~nm is selected to track the number density evolution of this species. To put the kinetic analysis (Section~\ref{kineticanalysis}) on a quantitative footing, we determine the oscillator strength of the 445~nm Py$^{\cdot+}$ band as follows. First the integrated absorbance of the 445~nm Py$^{\cdot+}$ band is plotted versus that of the 334~nm Py band during the course of VUV photolysis at different ice temperatures. These graphs are shown in Fig.~\ref{FIGURE2}. Note the tight, linear behavior between the loss of neutral pyrene and growth of the pyrene cation during early photolysis times up to 100~s (the first 10 successive datapoints). Inspection of Fig.~\ref{FIGURE2} shows that the slope is steepest and the ratio of the integrated absorbance of the Py$^{\cdot+}$ band to the Py band is optimum in the 25~K ice. Since no other photoproduct bands are evident during the linear correlation stage, we conclude that during this phase, neutral pyrene is solely converted to the cation as described previously for naphthalene and quaterrylene \citep{gudipati06a}. The straight line portion, fitted through the first 10 data points of irradiation at 25~K, is used to determine the oscillator strength of Py$^{\cdot+}$. Given that the ratio of the Py$^{\cdot+}$ to the 334~nm Py band is 0.99 and the oscillator strength of this Py transition is 0.33, the oscillator strength of the 443~nm Py$^{\cdot+}$ band in water ice is taken to be 0.33 as well. This conclusion is consistent with {\itshape ab initio} calculations on pyrene by \citet{weisman05}. They calculated that the oscillator strength of the cation is only $\sim$2\% stronger than that of the neutral species. 

As will be described below, the photolysis of Py in water ices at higher temperatures produces other species in addition to the cation. This explains the different curves in Fig.~\ref{FIGURE2}.

\subsection{HCO bands in Py:CO}
\label{sectionHCO}

\begin{figure}[t]
\centering
\includegraphics[width=8.5cm]{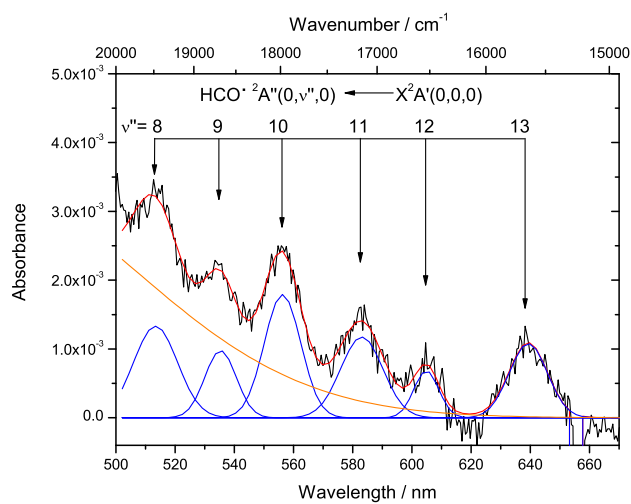}
\caption{Vibrational progression of HCO$^\cdot$ generated in a Py:CO ice at 10~K after 600 seconds of VUV irradiation plotted together with a Gaussian fit (red) to the absorption spectrum (black). The individual Gaussians are shown in blue. The orange line indicates the red wing of the underlying broad absorption feature (section~\ref{broadabsorption}). (This figure is available in color in electronic form.) }
\label{FIGURE3}
\end{figure}

VUV irradiation of a Py:CO ice also produces a vibrational progression ranging from $\sim$ 500 to 650~nm. As shown in Fig.~\ref{FIGURE3}, these absorptions, located at 513.4, 535.3, 556.3, 583.0, 604.9, and 639.2~nm, are assigned to the $^2{\rm A}''(0,\nu '',0)\leftarrow{\rm X}^2A'$(0,0,0) HCO$^\cdot$ ($\nu$''=8-13) transitions based on band positions reported by \citet{ijzendoorn83}. The clear HCO$^\cdot$ progression indicates a photolytic source of free H atoms in the ice. In addition, it confirms the ability of this setup to record reactive intermediates in the ice.

A possible explanation of the source of H atoms is related to the experimental conditions. The experiments reported here are performed under high vacuum (10$^{-7}$~mbar) conditions. Therefore, background H$_2$O vapor has ample time to condense onto the sample window while cooling down and while growing the ice sample. Water is well known to photodissociate upon VUV irradiation \citep[e.g.][]{oberg08a,andersson08}:
\begin{equation}
{\rm H_2O} \xrightarrow{VUV} {\rm OH^\cdot + H^\cdot}.
\end{equation}
An experiment on VUV irradiation of a \textquotedblleft pure\textquotedblright~CO ice indicated that HCO$^\cdot$ is also efficiently produced in the absence of pyrene. Therefore, it is likely that water contamination is responsible for the production of HCO$^\cdot$ via:
\begin{equation}
{\rm H^\cdot + CO \rightarrow HCO^\cdot}.
\end{equation}
Another possible formation route could be via VUV induced hydrogen abstraction from pyrene. This pyrene photodissociation reaction, however, is unlikely to occur, since PAHs are generally highly photostable molecules.

\subsection{The 400~nm band carrier}
\label{400nmband}

\begin{figure}[ht]
\centering
\includegraphics[width=9cm]{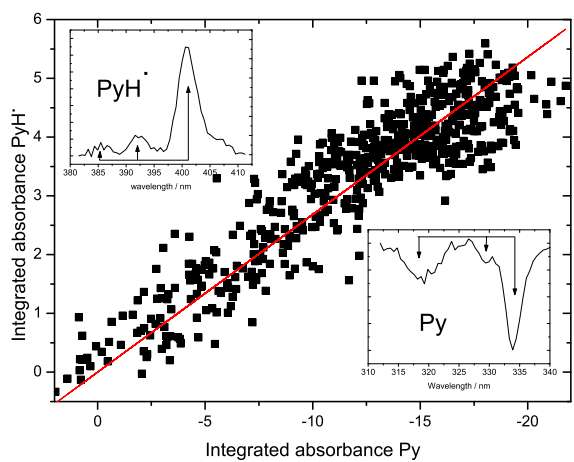}
\caption{Integrated absorbance of the 400~nm PyH$^\cdot$ band growth plotted against the loss of the 334~nm Py band {\itshape after} VUV irradiation is stopped in a at 10~K ice. The straight line directly reflects the relative oscillator strength of both bands as described in Section~\ref{cationabsorption}. The two insets show the PyH$^\cdot$ and Py vibrational progressions in a CO ice 90 minutes after photolysis is stopped.}
\label{FIGURE4}
\end{figure}

Another vibrational progression appears at 400.5, 392.5, and 378.4~nm in the CO ice experiments. As shown in Fig.~\ref{FIGURE4}, the 400.5~nm band dominates this progression. In contrast, a single band appears at 399.4~nm in the Py:\water ice upon VUV irradiation of the samples. The relative intensity of these bands varies with respect to the Py$^{\cdot+}$ bands. The 400~nm bands are more pronounced than the cation bands in the \water ice only at {\itshape high} temperatures, whereas they are more pronounced in the {\itshape low} temperature CO ice. 

Two additional measurements were performed to identify the carrier responsible for these transitions. A kinetic experiment was performed on non-VUV irradiated Py:CO ice. This ice showed no sign of pyrene ionisation by the Xe-lamp which is used as spectroscopic light source, nor of the production of HCO$^\cdot$ and of formation of the 400~nm band. Subsequently, the ice was irradiated with the VUV source for 10 minutes. The steady growth of the 400~nm band with VUV photolysis indicates that the species responsible for the 400~nm band is a product of VUV processing of the ice. Moreover, when the VUV irradiation is stopped, the 400~nm band carrier {\itshape continues} to grow at the expense of the remaining neutral pyrene. This indicates that the chemical reaction leading to the formation of this species is not directly photon dependent, but rather depends on the diffusion of a photoproduct. A similar experiment on a Py:\water ice at 25~K indicates that the same process also takes place in \water ice. The detection of HCO$^\cdot$ radicals in the ice and the inherent presence of free photolytic H atoms, hints that the growth of the vibrational progression starting at $\sim$400~nm could be the result of the reaction of pyrene with diffusing H atoms:
\begin{equation}
{\rm Py + H^\cdot~\rightarrow~PyH^\cdot}.
\label{eq6}
\end{equation}
This assignment to the 1-hydro-1-pyrenyl radical (PyH$^\cdot$) is supported by other experimental studies \citep{okada76,okada80}, where progressions at similar band positions are observed upon (laser) flash photolysis.

Contrary to the Py:\water experiments where pyrene is also efficiently ionized, the experiment on  PyH$^\cdot$ formation in CO shows no sign of other reaction products. The integrated absorbance of the growing PyH$^\cdot$ transition is plotted versus the integrated absorbance of the diminishing neutral in Fig.~\ref{FIGURE4}. Growth is tracked over a duration of more than 1.5 hours. Since there is a one-to-one conversion of Py to PyH$^\cdot$ in the Py:CO ice (Eq.~\ref{eq6}), as described in Section~\ref{cationabsorption}, we derive an oscillator strength of 0.089 for this species by fitting a line through the correlating absorbances in Fig.~\ref{FIGURE4}.

\subsection{The 405~nm band carrier}

Besides the Py$^{\cdot+}$ and the PyH$^\cdot$ bands, another distinct absorption is found in the spectra of VUV irradiated ices. This feature is located at 405.0~nm in the Py:\water and at 406.2~nm in the Py:CO experiment. In our previously published paper on low temperature Py:\water ice we tentatively assigned this absorption to a negative ion, Py$^{-}$ or PyO$^{-}$ \citep{bouwman09}. The experiments on Py:CO ices presented here enable us to rule this assignment out because of the nearly absent Py$^{\cdot+}$ transitions. Firstly, Py$^{-}$ is ruled out because a much stronger second Py$^-$ absorption band, expected at 490~nm \citep{montejano95}, is absent in our Py:CO experiment. Secondly, PyO$^-$ is also ruled out, because it should exhibit absorptions down to 350~nm \citep{milosavljevic02}, bands which are also absent in the Py:CO experiment. Additionally, in our previous paper we assumed that PyO$^{-}$ was a product of PyOH. The formation of PyO$^-$ is also unlikely in the absence of PyOH absorptions in these experiments, as discussed below.

The absorption at 405~nm does not correlate with the cation, nor with the PyH$^\cdot$ band. The band only appears during photolysis and hence is characterized as a VUV photon related product. From the literature it is known that a pyrene triplet-triplet ($^3A_{g}^-$$\leftarrow$$^3B_{2u}^+$) transition is expected at this wavelength upon laser excitation of pyrene in solution which populates the lowest member of the triplet manifold \citep[e.g.][]{hsiao92, langelaar70}. In order for the 405~nm band to arise from this triplet-triplet transition, the lowest level must be populated and remain so with a long enough lifetime to allow absorptions to the $^3A_{g}^-$ level. In the ice experiments reported here, there are a number of possible routes for pumping the $^3$Py state. The most obvious route is via a photo excitation followed by intersystem crossing:
\begin{equation}
{\rm ^1Py}~\xrightarrow{VUV}~{\rm ^1Py^*} \xrightarrow{isc} {\rm ^3Py}.
\label{Eqnpyreneisc}
\end{equation}
Triplet formation is found to decrease with decreasing temperature in an ethanol ice \citep{stevens67}. This translates to our experiment in a nearly absent 405~nm band in the low temperature Py:\water experiment, because of the high Py$^{\cdot+}$ formation efficiency. In the high temperature \water ice experiments, on the other hand, the 405~nm absorption is much stronger because pyrene is available.

In the CO ice, on the other hand, where Py$^{\cdot+}$ production is low, formation of the 405~nm band carrier appears to be very efficient at low temperatures. The production of the 405~nm band carrier needs VUV photons to be initiated. Again, pumping of the $^3$Py state can occur via Eq.~\ref{Eqnpyreneisc}. Moreover, CO has a dipole allowed electronic transition in the VUV. Hence, speculating, pumping of the $^3$Py state by collisional de-excitation of CO molecules exited by the VUV radiation provides a reaction path:
\begin{equation}
{\rm CO}~\xrightarrow{VUV}~{\rm CO^*},
\end{equation}
followed by:
\begin{equation}
{\rm CO^* + ^1Py} \rightarrow {\rm ^3Py} + {\rm CO}.
\end{equation}

In summary, while we cannot identify the carrier of the 405~nm band, the $^3A_{g}^-$$\leftarrow$$^3B_{2u}^+$ transition seems a plausible explanation.

\subsection{Broad absorption feature}
\label{broadabsorption}

Finally, we discuss a broad feature extending from about 350 to 570~nm that underlies the narrower bands reported in the previous sections that grows upon photolysis in all cases. This band probably arises from overlapping bands due to a number of Py/H$_2$O or Py/CO photoproducts. Part of this Py-residue feature remains even after warming up the sample window to room temperature, whereas all other features disappear at the water desorption temperature.

As discussed above, the very broad feature must be due to a variety of similar but distinct photoproducts, all containing the pyrene chromophore. Mass spectral analysis of the species produced by the VUV photolysis of a few other PAHs in water ice show that the parent PAH is not destroyed but that OH, O, and H are added to some of the edge carbon atoms \citep{bernstein99}. Given the multiplicity of the side sites on pyrene that can undergo substitution, it is likely that the photoproducts produced in the experiments reported here are multiply substituted, rather than singly substituted. Thus, it is plausible that a mixture of related but distinct Py-X$_n$ species, where X may be H, OH, or O, produce the broad band.

In our previous work we reported the production of a clear and reproducible PyOH band at 344.9~nm in a low temperature \water ice \citep{bouwman09}. The results presented here do not show this absorption feature. However, in some instances the absorption was found upon irradiation or warm-up of the ice. The irregular appearance of the PyOH absorption feature in these experiments indicates that the formation of this species is highly sensitive to the sample's physical parameters, {\emph i.e.} structure of the ice, temperature and concentration. One possible explanation is that in the previously reported experiments, the Py concentration was not controlled and those experiments sampled a very different ice concentration and, by implication, physical ice structure. While we do not have a solution for this discrepancy, we would like to stress that both measurement series have been fully reproducible over many independent experiments for periods of months. An experimental program to investigate the role of concentration on PAH:\water ice photochemistry is underway.

\section{Py:H$_2$O ice photochemistry at different temperatures}
\label{kineticanalysis}

\begin{figure}
\centering
\includegraphics[width=9cm]{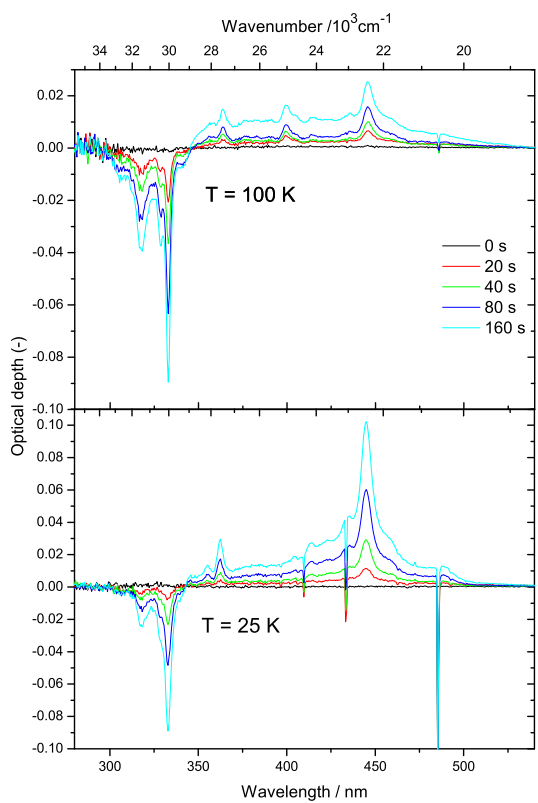}
\caption{The VUV induced spectroscopic changes in Py:H$_2$O ice for two different temperatures as a function of photolysis time. Comparing the spectra from the 25~K ice (bottom) with those of the 100~K ice (top) shows the critical role that temperature plays in determining  photochemical pathways in a PAH containing ice. In the 25~K ice, cation formation is favored over production of the pyrene residue and the 400 and 405~nm band carriers. The opposite holds for the 100~K ice.}
\label{FIGURE5}
\end{figure}
Figure~\ref{FIGURE5} shows the spectral evolution of two different Py:\water samples at different temperatures. The top frame presents the 280 to 540~nm spectra of the 100~K Py:H$_2$O ice after 0, 20, 40, 80 and 160~s of {\itshape in situ} photolysis and the bottom frame the corresponding spectra for the 25~K ice. These spectra are snapshots from the more than one thousand spectra collected during 4~hr of photolysis. They illustrate the rapid changes that take place during the early stages in the photochemistry of these ices and the major differences in reaction products at different temperatures.

To probe the VUV driven photophysics and reaction dynamics for a set of selected temperatures, the production and depletion of species was tracked as a function of irradiation time. To this end, the Py 334~nm, Py$^{\cdot+}$ 445~nm, and the PyH$^\cdot$ 400~nm bands were integrated for every spectrum. The spectra in Figs.~\ref{FIGURE1} and~\ref{FIGURE5} show that it is rather straightforward to determine the boundaries needed to integrate these bands. We estimate that the uncertainty in most of these band areas is of the order of 10\%.

\begin{figure*}[ht]
\centering
\includegraphics[width=14cm]{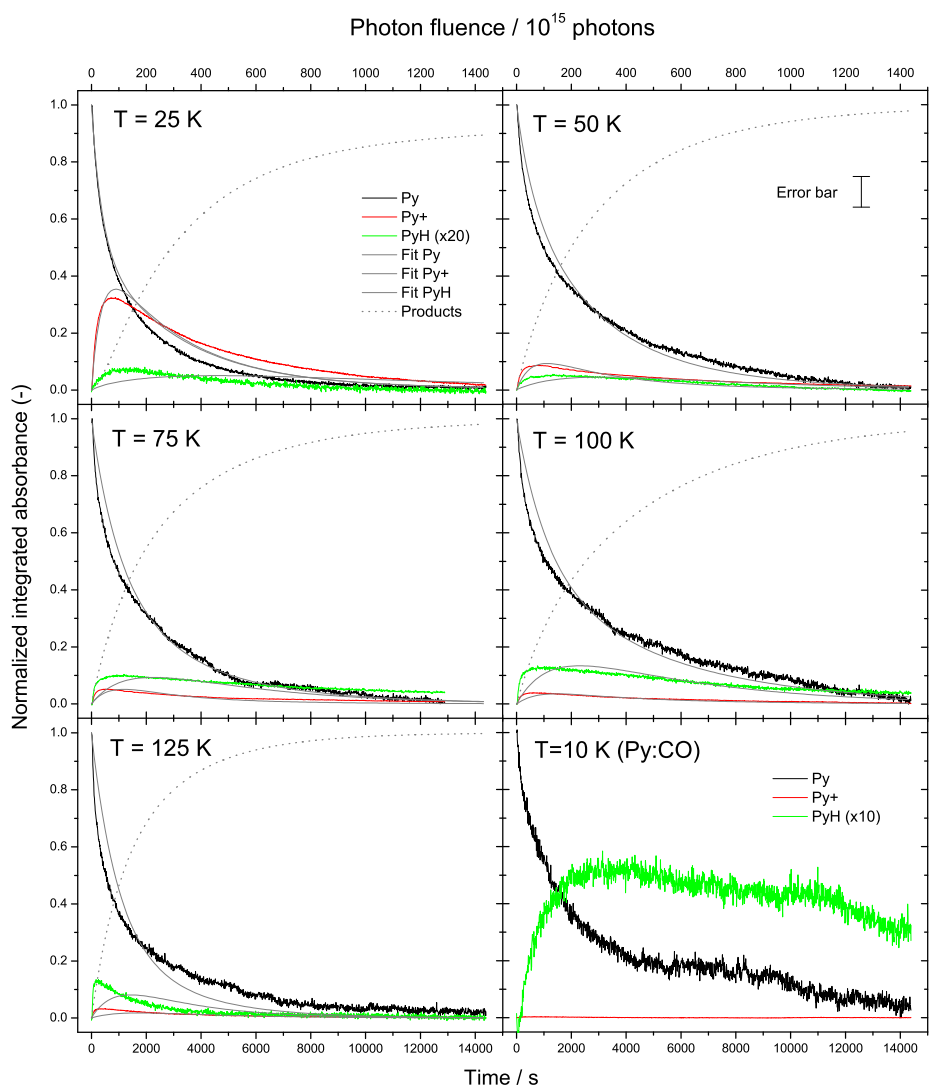}
\caption{The integrated absorbance of the Py 334~nm, Py$^{\cdot+}$ 445~nm, and PyH$^\cdot$ 400~nm bands as a function of VUV irradiation in Py:H$_2$O ices at 25, 50, 75, 100, and 125 ~K and a Py:CO ice at 10~K plotted together with the fits (grey lines) described in Section~\ref{kineticanalysis}. Integrated absorbance values are scaled and normalized to the initial value for the Py signal. The PyH$^\cdot$ feature is multiplied by a factor of 20 for the Py:H$_2$O experiments and by a factor of 10 for the Py:CO experiment. The approximate, overall growth of the total Py photoproduct band (P$_1$+P$_2$+P$_3$) is also shown (dotted line). (This figure is available in color in electronic form.)}
\label{FIGURE6}
\end{figure*}

The integrated absorbances of the neutral Py, strongest Py$^{\cdot+}$ and PyH$^\cdot$ bands in \water ice at temperatures of 25, 50, 75, 100 and 125~K are plotted versus photolysis time (VUV fluence) in Fig.~\ref{FIGURE6}. The spectra in Fig.~\ref{FIGURE5} and photochemical behavior in Fig.~\ref{FIGURE6} show that, upon photolysis, neutral pyrene loss is immediate and rapid. The initial growth of Py$^{\cdot+}$ mirrors the rapid, initial loss of Py. However, while Py steadily decreases, and several other Py photoproduct bands increase during some 4~hours of photolysis, the production of Py$^{\cdot+}$ reaches a maximum and then slowly diminishes. From Fig.~\ref{FIGURE6} one can clearly see that ionisation of pyrene is most efficient in the low temperature ice. Formation of PyH$^\cdot$, on the other hand, is much more efficient at higher temperatures.

For comparison, the integrated absorbances for the irradiated Py:CO ice are plotted as a function of time in the right bottom frame of Fig.~\ref{FIGURE6}. Note that the PyH$^\cdot$ band is multiplied by a factor of 10 in the Py:CO experiment, compared to a factor of 20 for the Py:\water experiment. The PyH$^\cdot$ band is clearly more prominent in the CO ice experiment compared to the \water ice experiments. The Py$^+$ signal on the other hand is negligible. This indicates that the \water ice plays a role in ion formation and stabilization.

To place this behavior on a quantitative footing, the integrated areas for the Py and Py$^{\cdot+}$ bands are converted to number densities using Eq.~\ref{eqns1}. Here, an oscillator strength of 0.33 is used for the 334~nm Py bands. The values used for the oscillator strengths of the Py$^{\cdot+}$ and PyH$^\cdot$ bands are 0.33 and 0.089, respectively, as determined in Sections~\ref{cationabsorption} and~\ref{400nmband}. Perusal of Fig.~\ref{FIGURE6} shows that Py behaves similarly in all of the ices considered here. Regardless of temperature, its signal drops quickly with the onset of irradiation and continues to diminish with ongoing photolysis. Likewise, Py$^{\cdot+}$ grows rapidly with initial photolysis but it peaks after a relatively short time interval corresponding to a fluence of roughly $8\times 10^{16}$ photons and then drops continuously. While the Py$^{\cdot+}$ growth and loss curves resemble one another, cation production efficiency is strongest in the 25~K ice. This efficiency remains of the same order at even lower temperatures (not shown here). The photolysis time required for the cation to peak shortens with temperature increase. The PyH$^\cdot$ band contribution is minor with respect to the Py$^{\cdot+}$ band for ices below 50~K. This reverses between 50 and 75~K, suggesting that there is a change in the dominant Py:H$_2$O ice photochemical channel in this temperature range.

\begin{figure}
\centering
\includegraphics[width=5cm]{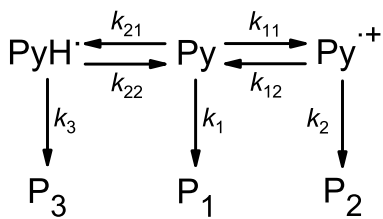}
\caption{Reaction scheme used to fit the experimental data.}
\label{FIGURE7}
\end{figure}

A kinetic analysis of the plots in Fig.~\ref{FIGURE6} is carried out using the reaction scheme indicated in Fig.~\ref{FIGURE7}. Here, $k_{11}$ is the photoionisation rate of Py to Py$^{\cdot+}$, $k_{12}$ the electron-ion recombination rate of Py$^{\cdot+}$, $k_{21}$ the production rate of the PyH$^\cdot$ feature, and $k_{22}$ the reverse reaction rate of PyH$^\cdot$ to Py. The rates designated $k_1$, $k_2$, and $k_3$ are the production rates for the different products that comprise the Py-residue band. The oscillator strengths for the Py$^{\cdot+}$ and PyH$^\cdot$ bands are also fitted, but are restricted to stay within $\pm$10\% of the experimentally determined values of 0.33 and 0.089. All reactions are assumed to be first order in the reactant. The relative abundances of \textquotedblleft free or solvated electrons\textquotedblright~and O, H and OH radicals in the ice are not considered.

\begin{figure}
\centering
\includegraphics[width=8.5cm]{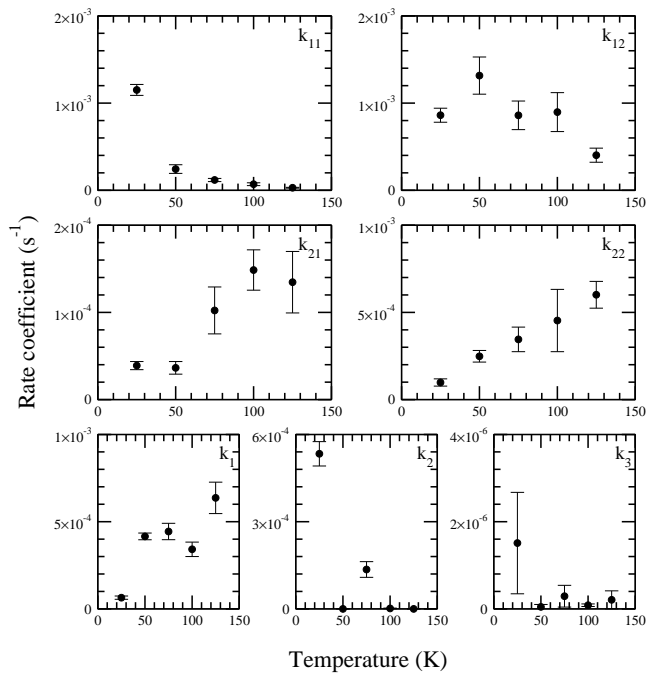}
\caption{Parameters ($k_{nm}$) as a function of temperature resulting from fitting the reaction scheme (Fig.~\ref{FIGURE7}) to the kinetic experiments (Fig.~\ref{FIGURE6}). All rates are indicated in s$^{-1}$.}
\label{FIGURE8}
\end{figure}

The fits to the growth and decay curves are included in Fig.~\ref{FIGURE6} and the temperature dependence of the derived rate constants is presented in Fig.~\ref{FIGURE8}. The agreement between the fit and the experimental data in terms of curve shape and absolute intensity is good. The fitted oscillator strengths of the Py$^{\cdot+}$ and PyH$^\cdot$ bands amount to 0.31 and 0.082, respectively, and hence do not deviate much from the experimentally determined values.

The graph in Fig.~\ref{FIGURE8} indicates that the Py photoionisation rate ($k_{11}$) drops rapidly between 25 and 50~K. The electron recombination rate ($k_{12}$) only decreases slightly, if at all, within the error over the entire temperature range. As mentioned above, the production of the PyH$^\cdot$ becomes more important at higher temperatures. Its formation rate ($k_{21}$) is small in all ices up to 50~K ($< 4.4\times 10^{-5}$), but jumps to $>1\times$10$^{-4}$ in the ices with temperatures of 75~K and higher. Also the back channel from PyH$^\cdot$ to Py, $k_{22}$, shows a temperature dependence. It increases nearly linearly in going from cold to warm ices. The formation rate of a photoproduct directly from Py ($k_1$) also seems to jump at 50~K. The formation rate of products from the Py$^{\cdot+}$ species, on the other hand, seems to lower with increasing temperature. Finally, the rate of product formation from the PyH$^\cdot$ channel is low throughout the entire temperature range. The jump in rate of the formation of P$_1$ and PyH$^\cdot$ with temperature probably reflects the diffusion barrier of radical species (H$^\cdot$ and OH$^\cdot$) in the ice.

Since published studies of the processes induced by the photolysis of other PAH:H$_2$O ices are limited, there is not much information available with which to compare these results. While, to the best of our knowledge, there are no reports on the photochemistry that takes place as a function of ice temperature or long term fluence, the VUV photochemistry of the PAHs naphthalene, 4-methylpyrene (4MP), and quatterrylene in water ice at 10~K has been studied \citep{gudipati03,gudipati04,gudipati06a,gudipati06}. Their results are in good agreement with the low temperature (25~K) case reported here. Namely, the parent PAH is easily and efficiently ionized, with quantitative conversion of the neutral species to the cation form. The focus in the earlier studies was on cation production and stabilization and not on long duration photolysis experiments. In their study of 4MP:H$_2$O (1:$>$500) ice at 15~K, \citet{gudipati03} utilized a reaction scheme similar to that on the right half of that presented in Fig.~\ref{FIGURE7}. Table~\ref{table2} compares the reaction rates they determined with those of the 25~K ice reported here. Except for the production of P$_2$, that deviates by one order of magnitude, there is very good agreement between the rate constants for each step in the two experiments.

The growth and decay curves in Fig.~\ref{FIGURE6}, taken together with the temperature dependence of the reaction rates in Fig.~\ref{FIGURE8}, show that the VUV driven PAH photochemistry depends strongly on ice temperature. The influence of the ice morphology on this chemistry was also investigated, to understand the origin of the temperature dependence. An experiment on an ice deposited at 25~K, annealed to 125~K, and subsequently cooled down to 25~K before photolysis, showed that the ionisation rate and efficiency are similar to that of an unannealed ice. Apparently, it is not the morphology but the temperature of the ice that primarily determines which process dominates. We discriminate between two temperature regimes. One governed by ion-mediated processes that dominate at 25~K and slightly higher temperatures and a second, presumably radical driven regime, that becomes increasingly more important at higher temperatures. 

\begin{table}
\begin{minipage}[b]{\columnwidth}
\caption{ The reaction rates for the VUV photolysis of Py:H$_2$O ($\sim$1:5,000)
ice at 25~K  compared to those for 4-methylypyrene:H$_2$O (1:$\leq$500) ice at 15~K \citep{gudipati03}. Photon rates are indicated for reaction channels which are dominated by photon processes.
\label{table2}} \centering \scriptsize
\begin{tabular}{l c r@{$\times$}l c c}
\hline
\hline
\multicolumn{2}{c}{Rate}&\multicolumn{2}{c}{This work}&Photon rate&Gudipati 2003\\
&&\multicolumn{2}{c}{(s$^{-1}$)}&(cm$^2$photon$^{-1}$)&(s$^{-1}$)\\
\hline
Py$\xrightarrow{VUV}$Py$^{\cdot+}$ &($k_{11}$)&(1.2$\pm$0.1)&10$^{-3}$&	1.2$\times$10$^{-17}$	&1.3$\times$10$^{-3}$\\
Py$^{\cdot+}$+e$^{-}\rightarrow$Py & ($k_{12}$)&(9$\pm$2)&10$^{-4}$&	9$\times$10$^{-18}$	&8$\times$10$^{-4}$\\
Py $\rightarrow$ P$_1$& ($k_1$)&(5$\pm$1)&10$^{-5}$&...&4$\times$10$^{-5}$\\
Py$^{\cdot+}$ $\rightarrow$ P$_2$& ($k_2$)&(5$\pm$1)&10$^{-4}$&...
&5$\times$10$^{-5}$\\
\hline
\end{tabular}
\end{minipage}
\end{table}

\section{Astrochemical Implications}
\label{sectionAstrochemical}

As shown in the previous sections, ionisation and chemistry of a rather small PAH, pyrene, trapped in \water ice turns out to be very efficient in a laboratory setting. Here, we extend these findings to interstellar conditions, with the aim to include rates in astrochemical models. For this, it is crucial to disentangle pure photochemical processes from diffusion, since the latter will be highly dependent on the number density of radicals and electrons in the ice. As mentioned in the previous section, the photoionisation of Py is probably a single photon process, whereas protonation of Py and the electron recombination of Py$^{\cdot+}$ are results of both VUV photolysis and diffusion. The mechanism for PyH$^\cdot$ deprotonation is not clear, since it can either proceed through VUV processing or through hydrogen abstraction by diffusing species. Diffusion of radicals through the ice is a thermally activated process and will therefore increase with temperature. Recombination, however, is largely temperature independent in our experiments, indicating that the rate of Py$^{\cdot+}$ recombination is not dominated by diffusion of electrons in the ice. If Py$^{\cdot+}$ loss is via electron recombination and not Py$^{\cdot+}$ reaction with H$_2$O or one of its photoproducts, the electron most likely originates from the initial photoionisation event where electrons stay in the vicinity of the recombining Py$^{\cdot+}$ species. Hence, this local process can be, although indirectly, regarded as a single photon process. 

The rates of protonation of Py and deprotonation of PyH$^\cdot$ show a temperature dependence and the importance of diffusion can therefore not be excluded. This makes it harder to directly translate the rates (s$^{-1}$) to photon rates (cm$^{2}$photon$^{-1}$). However, we can determine astrochemical photon rates for both ionisation and recombination of pyrene in interstellar \water ice (see Table~\ref{table2}).

Now, to translate this to the astrochemical situation and compare this to other processes, let us assume that PAHs generally have ionisation rates similar to that of pyrene. How do ionisation and chemistry compare with other processes such as photodesorption of the icy grain mantle, in which the PAHs are embedded? To exemplify this, the rate of ionisation of {\bf a PAH in water ice at 25~K (in photon$^{-1}$) is calculated anywhere in a dense cloud where $A_{V}=3$} and compared to the VUV photodesorption rate of \water as derived by \citet{oberg08a}. {\bf It is well established that the onset of ice formation occurs in clouds with an edge-to-edge (through the cloud) magnitude of $A_V=3$ \citep[e.g.][]{whittet01}. Thus inside our hypothetical dense cloud at $A_V=3$ (from cloud edge to within the cloud) ices are present.}

The experimentally determined PAH ionisation rate in \water at 25~K, normalized to the total amount of deposited PAH is given by:
\begin{equation}
{\rm k}_{11}=\frac{{\rm d}\frac{[{\rm PAH}^+]}{[{\rm PAH}]_0}}{{\rm d}t}= 10^{-3}~{\rm s}^{-1}.
\end{equation}
Consider a typical interstellar grain, covered by a 100~monolayer (ML) thick ice. The number of sites on a grain is 10$^{15}$~cm$^{-2}$. If we assume that one in every 10$^4$ particles on the grain is a PAH, the total number of PAH molecules on the grain is $[{\rm PAH}]_0=100\cdot 10^{15}\cdot 10^{-4}=10^{13}~{\rm cm}^{-2}$. Furthermore, the  VUV photon flux in our laboratory, $\Phi$, is $10^{14}~{\rm photons~cm}^{-2}{\rm s}^{-1}$. The production rate of PAH cations on an interstellar grain is now given by $[{\rm PAH}]_0\cdot k_{11}/\Phi=10^{-4}~{\rm photon}^{-1}$. This ionisation rate is an order of magnitude lower than the rate of photodesorption ($\sim 10^{-3}~{\rm photon}^{-1}$) \citep{oberg08a}.

However, in {\bf our dense cloud} the number of photons available for PAH photoionisation is larger than the number of photons available for photodesorption of \water ice. This is because \water photodesorption occurs primarily by VUV photons, whereas PAH ionisation can occur with much lower energy photons. To quantify the radiation field in a {\bf dense }cloud {\bf at} $A_{V}=3$ as a function of wavelength ($\lambda$), we take the average UV interstellar radiation field ($I_\nu$) from \citet{sternberg88} and rewrite the expression to I$_\lambda$ with units photons~cm$^{-2}$s$^{-1}$nm$^{-1}$:
\begin{equation}
I_\lambda=\frac{1.068\times 10^{-4}c}{\lambda^3}-\frac{1.719\times 10^{-2}c}{\lambda^4}+\frac{6.853\times 10^{-1}c}{\lambda^5},
\end{equation}
where $c$ is the speed of light in nm~s$^{-1}$. The attenuation of the radiation field by dust as a function of wavelength is given by:{\bf
\begin{equation}
D_\lambda = \exp(\frac{-11.6A_V}{R_V}\frac{A_\lambda}{A_{1000\text{\AA}}}),
\end{equation}
from \citet{draine96}, where we assume $R_V=3.1$ and $A_V/A_{1000\text{\AA}} = 0.21$ \citep{whittet03}. This results in:}
\begin{equation}
D_\lambda=\exp(-0.8\frac{A_\lambda}{A_V}A_V),
\end{equation}
where the table of $A_\lambda/A_V$ values is taken from \citet{mathis90}. The photon flux per second per wavelength interval is given by:
\begin{equation}
P_\lambda=I_\lambda D_\lambda.
\end{equation}

Water ice absorbs photons with wavelengths ranging from 130 to 150~nm \citep{kobayashi83,andersson08}. The ionisation energy of PAHs on the other hand, is lowered by {\bf about 2~eV when in} \water ice \citep{gudipati04a,woon04}. For the wavelength range available for ionisation of PAHs, assuming that \water blocks all photons below 150~nm, we take 150 to 250~nm \citep{li01}. Integrating the photon flux in a cloud of $A_{V}=3$ over both wavelength intervals gives a number of photons available for PAH ionisation which is 6 times larger than the number of photons available for photodesorption of \water. Additionally, {\bf at $A_V=3$}, the cosmic ray induced UV field is negligible compared to the interstellar UV field \citep{shen04}. Therefore, the occurrence of photoionisation is of similar order as photodesorption of the main component in the grain mantle in a {\bf dense} cloud. The ionisation rates from Table~\ref{table2} can be directly included in astrochemical models in the form:
\begin{equation}
\frac{{\rm d[PAH^+]}}{{\rm d}t}=k_{11}\Psi{\rm[PAH]},
\end{equation}
where [PAH$^+$] is the concentration of the PAH (pyrene) cation in the ice, $k_{11}$ is the photon rate in cm$^2$photon$^{-1}$, $\Psi$ is the photon flux in photon~s$^{-1}$cm$^2$, and [PAH] is the PAH (pyrene) concentration in the ice.

In the above calculation we assume that all PAHs exhibit the ionisation behavior of the pyrene chromophore. Of course, more PAHs need to be investigated experimentally in order to draw conclusions on their general photochemical behavior in interstellar ices. However, if all PAHs have ionisation rates similar to pyrene, photoionisation and subsequent chemical reactions of PAHs trapped in ices are important processes in dense clouds. When frozen out in ices, PAHs have an important impact on the radical and electron budget in solid state chemistry. Hence, the processes described here may be more important than previously assumed in modeling complex interstellar grain chemistry. 

\section{Conclusions}
\label{conclusions}

A recently constructed setup has been used to track, on a sub-second timescale, the photochemistry of a PAH in \water and CO ices as a function of temperature. The setup used here clearly has advantages compared to the relatively slow infrared photochemical ice studies. The conclusions from this work on the PAH pyrene trapped in an interstellar ice analogue are summarized below: 

\begin{enumerate}
\item	A set of photochemical reaction products has been assigned, both in irradiated Py:\water and Py:CO ice experiments. The reaction products result from direct photoionisation of pyrene, or from a reaction of the parent, pyrene, with free H atoms produced in the matrix. Additionally, an absorption band is tentatively assigned to a triplet-triplet transition of pyrene. A vibrational progression assigned to HCO$^\cdot$ is found in spectra of the VUV irradiated Py:CO ice.    
\item Pyrene is easily and efficiently ionised when trapped in \water ice. Photoionisation is a non-diffusion related reaction and hence a photon-rate of $1.2\times10^{-17}$~cm$^2$~photon$^{-1}$, which can serve as input for astrochemical models, is derived.
\item When trapped in CO ice, pyrene ionisation is inefficient compared to water ice.
\item	Electron-ion recombination is independent of ice temperature and is characterized as a non diffusion dominated reaction. For this process, a photon rate of $9\times10^{-18}$~cm$^2$~photon$^{-1}$ is derived which can be directly used in astrochemical models. 
\item There are two distinct reaction paths in the photochemistry of pyrene trapped in \water ice. At low temperatures ($<50$~K) the chemistry is dominated by ion-molecule interactions and processes. At temperatures above 50~K reactions are dominated by diffusing radical species.
\item A simple model indicates that, in {\bf dense clouds where $A_V=3$}, the rate of pyrene ionisation is comparable to the rate of photodesorption in water rich ices. Hence, chemical reactions involving pyrene, and its cation can be important in modeling grain chemistry in such environments.
\end{enumerate}

\begin{acknowledgements}

This work is financially supported by \textquoteleft Stichting voor Fundamenteel Onderzoek der Materie\textquoteright (FOM), \textquoteleft the Netherlands Research School for Astronomy\textquoteright (NOVA) and NASA's Laboratory Astrophysics and Astrobiology Programs.  L.~J. Allamandola is especially grateful to the \textquoteleft Nederlandse Organisatie voor Wetenschappelijk Onderzoek\textquoteright (NWO) for a visitors grant.

\end{acknowledgements}

\end{document}